\documentclass[
final            
  ]
  {aipproc}

\usepackage[letterpaper]{geometry}
\layoutstyle{8x11single}
\usepackage{setspace}
\begin{document}

\title{Development and Validation of the Colorado Learning Attitudes about Science Survey for Experimental Physics}


\author{Benjamin M. Zwickl}{
	address={Department of Physics, University of Colorado Boulder, Boulder, CO 80309}
}
\author{Noah Finkelstein}{
	address={Department of Physics, University of Colorado Boulder, Boulder, CO 80309}
}
\author{H. J. Lewandowski}{
	address={Department of Physics, University of Colorado Boulder, Boulder, CO 80309},
	altaddress={JILA, University of Colorado Boulder, Boulder, CO 80309}
}

\begin{abstract}
As part of a comprehensive effort to transform our undergraduate physics laboratories and evaluate the impacts of these efforts, we have developed the Colorado Learning Attitudes about Science Survey for Experimental Physics (E-CLASS). The E-CLASS assesses the changes in students' attitudes about a variety of scientific laboratory practices before and after a lab course and compares attitudes with  perceptions of the course grading requirements and laboratory practices.  The E-CLASS is designed to give researchers insight into students' attitudes and also to provide actionable evidence to instructors looking for feedback on their courses.  We present the development, validation, and preliminary results from the initial implementation of the survey in three undergraduate physics lab courses. 
\end{abstract}

\maketitle


\section{Introduction}

There is a national need to evaluate lab courses, fueled in part by a growing desire to transform traditional laboratory experiences in science, technology, engineering, and mathematics (STEM) education \cite{President'sCouncilofAdvisorsonScienceandTechnology2012}.  In physics, a traditional experience usually begins with following procedural instructions in order to demonstrate well-understood phenomena and/or replicate historic experiments.  Toward the end of a traditional lab activity, it is common to have students perform error analysis calculations to compare their data with a well-known result.  Finally, the traditional activity is capped off with a written lab report.  In this lab experience, most of the intellectual work takes place after the data are collected, rather than during the process of designing and carrying out the experiment.  Additionally, traditional labs have focused on physics content and gaining familiarity with lab equipment.  The limited scope of these goals, along with concerns about the pedagogical effectiveness of labs, has prompted the development of a variety of redesigned lab curricula, especially for the introductory level.

Many redesigned labs seek to create increasingly authentic experiences with experimental science, which may increase retention in STEM degree programs \cite{Laursen2010b, Fortenberry2007} and as called for nationally \cite{President'sCouncilofAdvisorsonScienceandTechnology2012}.  Within the physical sciences, non-traditional transformed labs have emphasized a variety of goals: conceptual understanding of measurement and experimental design \cite{Buffler2008a, Kung2005},  microcomputer-based labs for building models and improving conceptual understanding \cite{Redish1997}, computational modeling \cite{Buffler2008}, inquiry-oriented labs \cite{Weaver2008}, and research experiences \cite{Horsch2012}.

Meanwhile, attempts to assess college-level physics labs include: conceptual gains \cite{Redish1997}, measurement concepts \cite{Buffler2008a, Kung2005}, data analysis skills \cite{Day2011}, and scientific practices \cite{Etkina2006b,Etkina2010}.  However, what is lacking is a common evaluation tool that can be applied to a variety of lab experiences.  One particular aspect of the lab that can be evaluated easily, and which also aligns with a common learning goal, is the development of students' attitudes, beliefs, and practices regarding experimental physics.  

There has already been extensive work looking at student attitudes and beliefs about knowing and learning physics, including: the Views of Nature Science Questionnaire (VNOS) \cite{Lederman2002}, the Maryland Physics Expectations Survey (MPEX) \cite{Redish1998}, the Epistemological Beliefs Assessment for Physical Science (EBAPS) \cite{Elby2001}, and the Colorado Learning Attitudes about Science Survey (CLASS) \cite{Adams2006}.  The primary use for these surveys has been in large introductory physics courses.

The Colorado Learning Attitudes about Science Survey for Experimental Physics \cite{website:E-CLASS1140_Short} differs from previous surveys in several important ways.  First, knowing and learning experimental physics encompasses a different set of goals and ideas than most lecture-based courses.  The E-CLASS was specifically designed to assess the impact of \emph{lab} courses on students' attitudes and beliefs about \emph{experimental} physics.   Second, we want to evaluate the longitudinal impact of a lab sequence, so the survey is useful for evaluating students at all stages of an undergraduate program.  Third, the survey compares students' beliefs with their practices.  Lastly, the survey is designed to guide and assess lab course transformation by giving instructors actionable evidence about their lab course.

\vspace*{-10pt}
\section{Design and Structure}

In order to design a tool that can assess the attitudes, beliefs, and practices of students at a wide range of levels, we required a definition of authentic experimental physics.  This definition was furnished by a set of consensus learning goals, developed through significant involvement of the University of Colorado Boulder (CU) Physics faculty as part of a course transformation of our senior-level advanced lab course \cite{Zwickl2011}.  Most E-CLASS question categories were based on these learning goals, but some came from existing attitudes and beliefs surveys. Table \ref{tab:Question_categories} lists the major categories for E-CLASS questions.  These are not categories in the sense of some earlier surveys where we expect a high degree of correlation among statements within a category, but are groups of questions surrounding a common theme.  For instance, two statements in the \emph{Physics Community} category are ``I am able to read a scientific journal article for understanding.'' and ``Communicating scientific results to peers is a valuable part of doing physics experiments.''  Both of these statements express how experimental physics extends beyond the individual researcher; however they clearly represent distinct practices, and any particular lab course could emphasize them to varying degrees.

\begin{table}[t]
    \caption{List of E-CLASS assessment categories.  Italics denotes upper-division lab learning goals.}
    \begin{tabular}{p{0.2\textwidth}  p{0.2\textwidth}}
        \hline

        Affect & \em{Argumentation} \\
        Confidence & \em{Experimental design} \\ 
        \em{Math-Physics-Data connection} &   \em{Modeling the measurement system} \\
        \em{Physics community} & Purpose of labs \\
        \em{Statistical uncertainty} & \em{Systematic error} \\
        \em{Troubleshooting} & \\	
       \hline
    \end{tabular}
    \label{tab:Question_categories}
\end{table}

\begin{figure}
\includegraphics[width=0.45\textwidth, clip, trim=0mm 0mm 0mm 0mm]{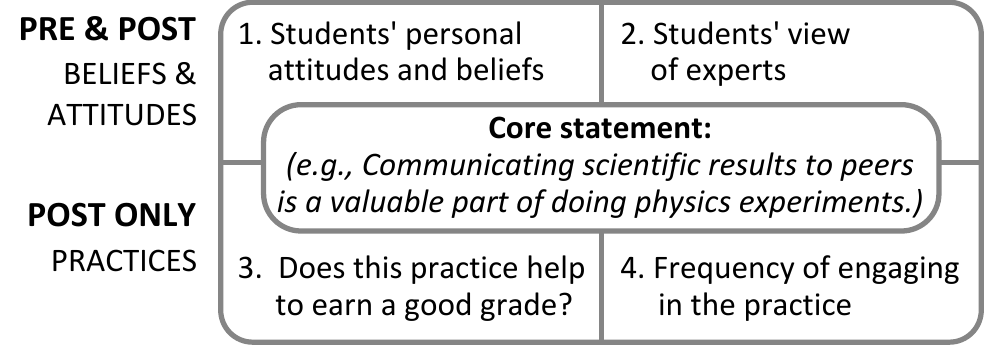}
\caption{The E-CLASS assesses each core statement in four ways.  Category: Physics community.}\label{fig:Four_quadrants}
\end{figure}

\begin{figure}
\includegraphics[width=0.45\textwidth, clip, trim=0mm 0mm 0mm 0mm]{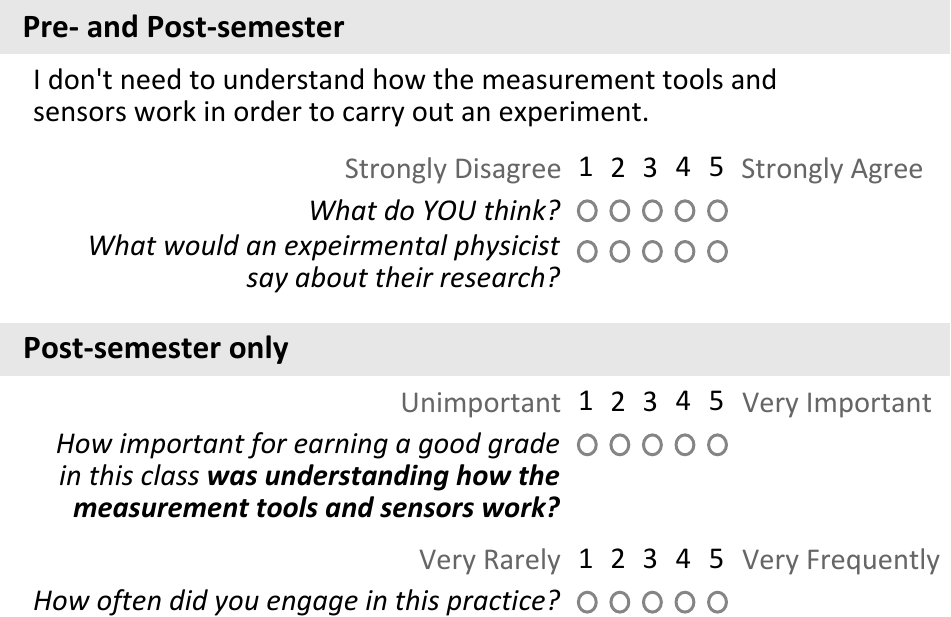}
\caption{A set of four questions from the E-CLASS Survey surrounding one core experimental practice: understanding the measurement probes and sensors. Category: Modeling the measurement system.}\label{fig:Example_question}
\end{figure}

One distinctive feature of the E-CLASS is that it evaluates students in four different aspects of their attitudes and experience with experimental physics, both generally, and specifically about the course.  The four quadrants of assessment around a single core idea are shown in Fig.~\ref{fig:Four_quadrants}.  The first two quadrants, which ask about students' personal attitudes and their beliefs about what experts would think \cite{McCaskey2004, Gray2008a}, are assessed at the beginning and end of the semester.  As shown in Fig.~\ref{fig:Example_question}, the personal and expert beliefs are presented adjacently and ranked on the same scale. The third and fourth quadrants relate to students' impression of the course.  Specifically, students are asked how important a particular practice was for earning a good grade and how often they engaged in that practice. The lower quadrants are assessed only at the end of the semester because they depend on students' experience in the class.  These questions give instructors insight into how the course structure influences students' attitudes and beliefs.

 In total, there are 23 core statements that are evaluated in all 4 quadrants and seven additional statements related to affect, confidence, and the purpose of labs, which are assessed only using the personal/expert pairs.  There are 60 Likert scale items on the pre-test and 106 Likert scale items in the post-test.  Because the questions all come in pairs and have similar wording, students are able to answer them quickly.  Most students take 10-15 minutes to complete the end-of-semester survey.

\vspace*{-10pt}
\section{Validation}

Validation of the survey took part in two ways.  First, students were given the survey in an interview format in order to ensure that students could consistently interpret questions in a way that aligned with our intended meaning and that we could accurately interpret their responses.  Second, physics faculty completed the survey to ensure that there was a consistent expert opinion. 

Student validation began with ten students who were interviewed while taking an early version of the E-CLASS.  These interviews resulted in substantial revisions of the survey, some of which are detailed below.  A revised version of the survey presented here was validated through nine additional  interviews in which students first took the survey by themselves and then were asked to explain their answer on each question.  During the interview, students were also asked to define technical terms such as ``approximation'' and ``systematic error'' in their own language so that questions were phrased in a way that would make sense to students coming into the introductory physics lab.  This population of nine student interviewees included four female and five male students and included two students prior to taking any college physics labs, four students in the intro lab, and three students in the senior-level advanced lab.

One significant change that came about through student interviews was the choice to change from a single statement asking about the student's personal beliefs to the personal/expert pairs  shown in Fig. \ref{fig:Example_question}.  Many aspects of lecture courses, such as textbooks and homework problems, have no parallel in professional practice.  However, terms like ``lab'' and ``experiment'' are used equally to describe lab courses and professional research, so there was an inherent ambiguity when asking a question about students beliefs about experimental physics.  Is the question asking about what I personally experience/believe or what professional researchers do?  Is the question asking about \emph{all} physicists, or just experimental physicists? When these ambiguities surfaced in the interviews we evolved the wording into the present form contrasting \emph{``What do YOU think?''} with \emph{``What would an experimental physicist would say about their research?''}

Six faculty in experimental physics completed the survey to determine if there was consensus on what ``an experimental physicist would say about their research.''  All six faculty had matching agree or disagree responses for 22 of the 30 statements.  Five of the 30 statements had five matching responses and one neutral.  Two of the 30  statements had four matching responses: ``Working in a group is an important part of doing physics experiments,'' and ``When I encounter difficulties in the lab, my first step is to ask an expert, like the instructor.''  Despite not achieving perfect consensus, these questions have majority consensus and probe important aspects of experimental research such as collaboration and independence in problem solving.  The final statement, which had the least consensus (2 Agree, 3 Neutral, 1 Disagree), was ``Nearly all students are capable of doing a physics experiment if they work at it.''  When reflecting on their professional research, it is not surprising many faculty think some students aren't capable since highly trained graduate students often struggle.

\vspace*{-10pt}
\section{Results}

The first set of E-CLASS data, which are presented here, were collected at the end of the Spring 2012 semester at CU.  The survey was administered in the entire physics lab sequence ($N$ denotes students that completed the survey): Introductory calculus-based physics lab ($N=242$), Sophomore-level Modern Physics Lab ($N=45$), Junior-level Electronics Lab ($N=3$), and the Senior-level Advanced Lab ($N=16$).  Since the response from the Junior-level electronics lab was so small, the results are omitted from the analysis.

With 106 Likert scale questions, there are many possible research questions we can investigate.  Here we choose to look at the four quadrants surrounding the particular core statement "Doing error analysis (such as calculating the propagated error) usually helps me understand my results better.   This statement is interesting because of its prominence in traditional physics labs, including our introductory lab at CU.

Much of the content of error analysis is summarized in the commonly used textbook by Taylor \cite{Taylor1997}.  At CU, where these survey data were generated, the introductory physics lab includes six lectures, all of which are devoted to aspects of measurement, uncertainty, and error analysis.  Also, every introductory lab experiment requires a lab report with a specific section on error analysis.  The heavy emphasis on error analysis is motivated by the belief of experimental physicists that a numerical quantity without an estimate of uncertainty is not useful.  

\begin{figure}
\includegraphics[width=0.43\textwidth, clip, trim=0mm 0mm 0mm 0mm]{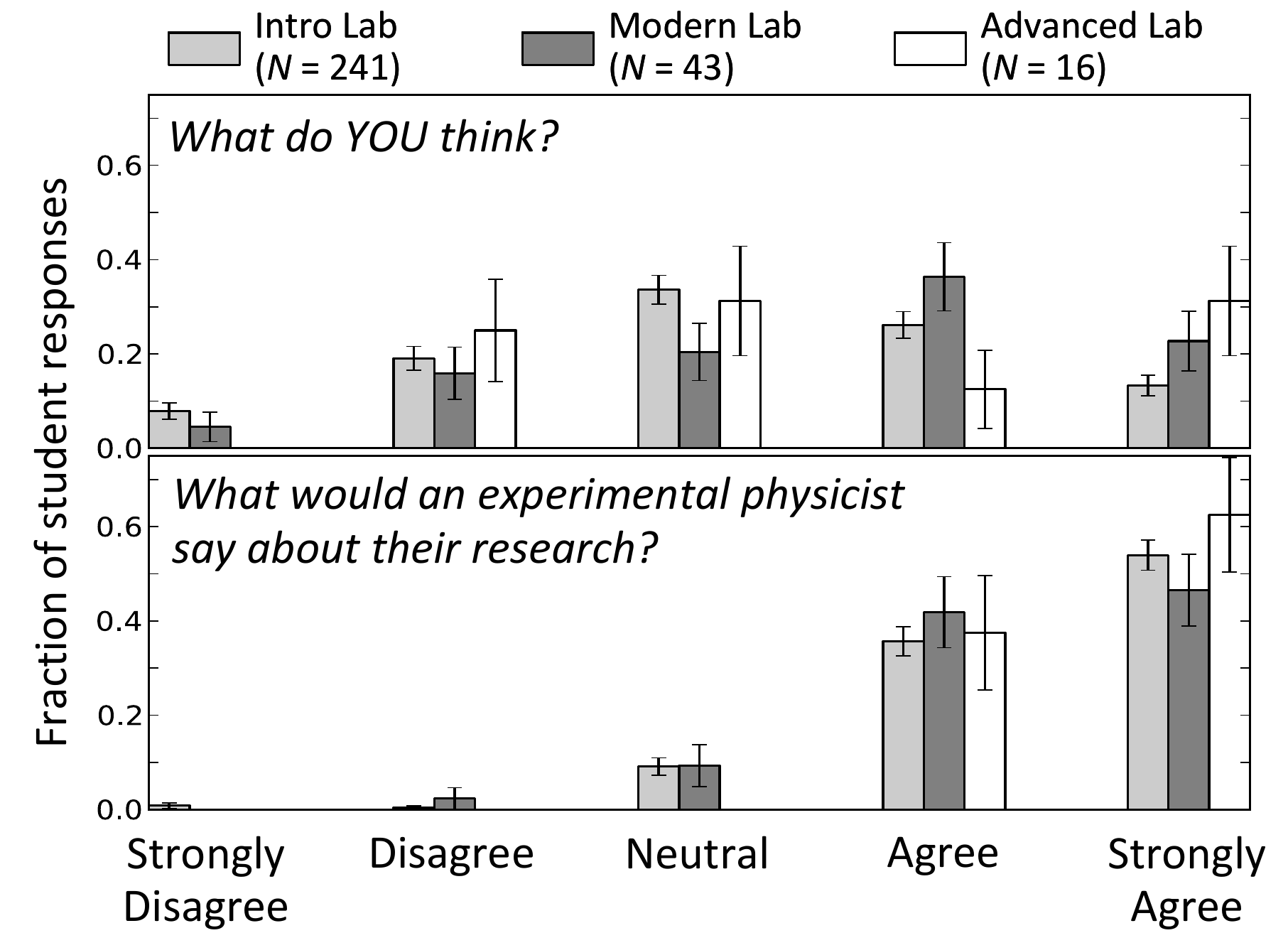}
\caption{Comparison of students' attitudes in three laboratory courses regarding the statement ``Doing error analysis (such as calculating the propagated error) usually helps me understand my results better.''  Category: Statistical uncertainty.}\label{fig:Error_Analysis}
\end{figure}

With all this emphasis, do students find that error analysis usually helps them better understand their results?  Fig. \ref{fig:Error_Analysis} shows histograms of students' responses to this statement from the three different levels of physics courses spanning the introductory to the advanced lab course.  A clear pattern is that while students realize experimental physicists find error analysis a valuable tool for understanding data (in fact, all 6 faculty agreed), many students' are personally neutral or disagree that error analysis helps them understand their data.  Also, this result holds for many students completing higher level lab classes despite the fact these students are more likely to be physics majors who have used error analysis in their previous lab courses and research experiences.

\begin{figure}
\includegraphics[width=0.43\textwidth, clip, trim=0mm 0mm 0mm 0mm]{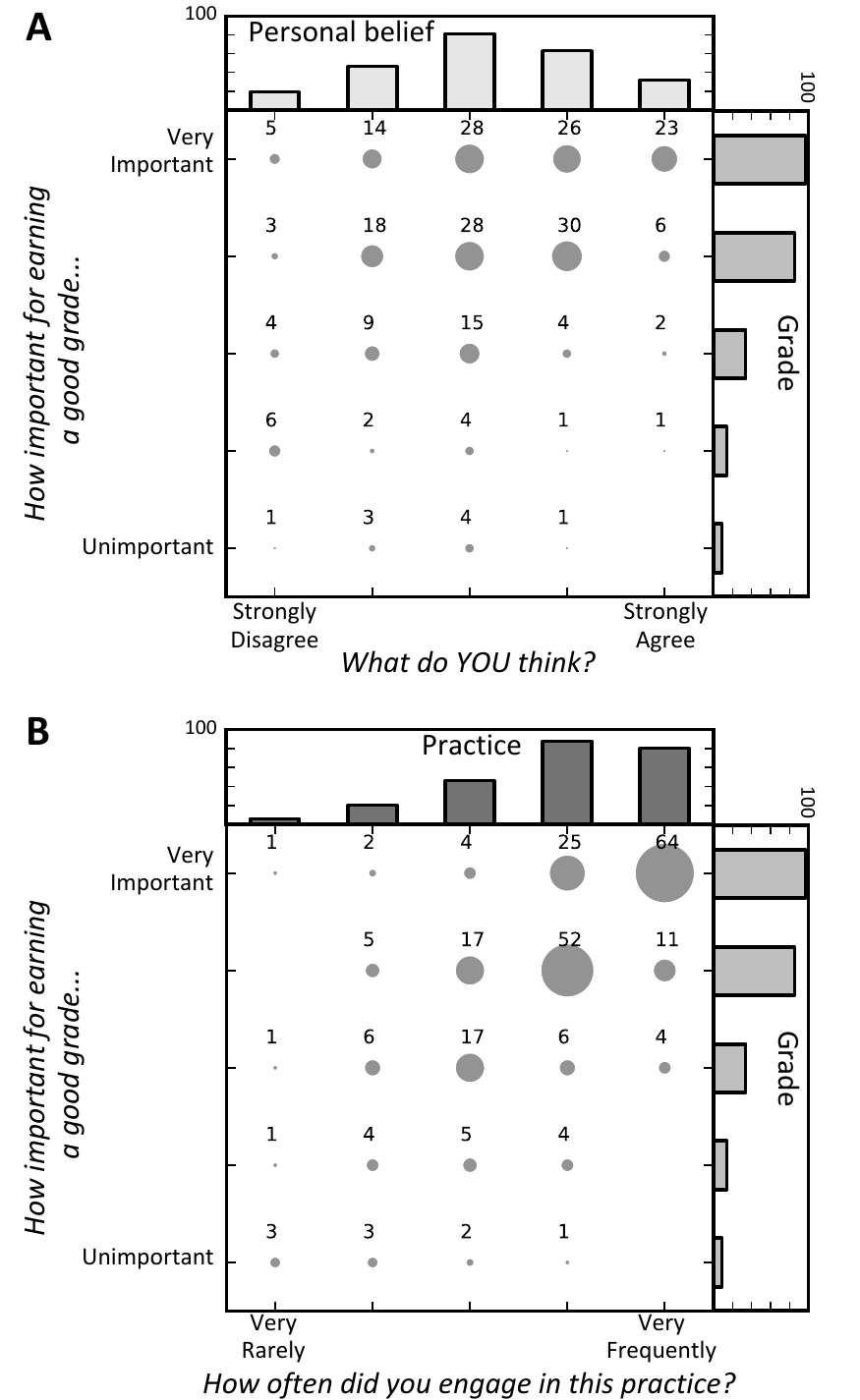}
\caption{Comparison of three quadrants (personal belief, grade, and practice) around the idea of error analysis. The numerical values are number of students.  The bar graphs represent the total students in a column or a row.}\label{fig:Error_Analysis_Scatter}
\end{figure}

One possible explanation for students' less-than-expert-like personal belief about helpfulness of error analysis is that courses did not have sufficient grade incentive or opportunity to engage in the routine practice of using error analysis in the lab.  However, Fig. \ref{fig:Error_Analysis_Scatter} shows this is not the case.  Fig. \ref{fig:Error_Analysis_Scatter} shows the correlation between introductory lab students' personal beliefs, and their responses to the questions \emph{``How important for earning a good grade in this class was using error analysis (such as calculating the propagated error) to better understand your data?''} and \emph{``How frequently did you engage in this practice?'}'  Fig. \ref{fig:Error_Analysis_Scatter}A  shows that students think error analysis is important for earning a good grade in the class, and Fig. \ref{fig:Error_Analysis_Scatter}B shows that they practice error analysis regularly (in fact, as part of every lab).  Despite this emphasis, many students do not find it helpful as a tool for understanding.

So, although our grading and practices in the introductory lab are aligned with a major learning goal for the course, students' experiences with error analysis are at best only partially successful in demonstrating the utility of error analysis.  This finding agrees with the common instructor experience that students treat error analysis as an exercise in following algorithms rather than in sense-making. The error analysis does not aid in interpreting or redesigning the experiment, or making a decision.  The results of the error analysis are never used for any purpose beyond a comparison with a  known result.  This approach differs from some transformed lab experiences that focus on conceptual understanding of measurement and uncertainty \cite{Kung2005}.

Although these results focus on just one of the 30 main concepts and ideas, they give a picture of how the E-CLASS can be used to probe students' attitudes and beliefs about experimental physics and to learn about how students' practices and the course structure influence these beliefs.

In conclusion, the Colorado Learning Attitudes about Science Survey for Experimental Physics has been designed and preliminarily validated.  This tool gives us a window into both students' views about experimental physics and about the nature of the lab courses they are taking.  Future semesters will provide an opportunity to administer the survey at the beginning and the end of the semester to measure shifts over the course of the semester.  We are hoping to administer the E-CLASS at other institutions during Fall 2012 in order to investigate the impact of a variety of transformed  lab courses.

This work was supported by National Science Foundation TUES award DUE-1043028.





\bibliographystyle{aipproc}   

\vspace*{-10pt}
\bibliography{Publications-PERC2012_MOD_et_al}

\begin{thebibliography}{21}
\expandafter\ifx\csname natexlab\endcsname\relax\def\natexlab#1{#1}\fi
\providecommand{\enquote}[1]{``#1''}
\expandafter\ifx\csname url\endcsname\relax
  \def\url#1{\texttt{#1}}\fi
\expandafter\ifx\csname urlprefix\endcsname\relax\def\urlprefix{URL }\fi
\providecommand{\eprint}[2][]{\url{#2}}

\bibitem[{PCAST}(2012)]{President'sCouncilofAdvisorsonScienceandTechnology2012}
{PCAST}, {Engage to Excel}, Tech. rep. (2012).

\bibitem[Laursen et~al.(2010)]{Laursen2010b}
S.~Laursen, et~al., \emph{{Undergraduate Research in the Sciences: Engaging
  Students in Real Science}}, Jossey-Bass, 2010.

\bibitem[Fortenberry et~al.(2007)]{Fortenberry2007}
N.~L. Fortenberry, et~al., \emph{Science} \textbf{317}, 1175--6 (2007).

\bibitem[Buffler et~al.(2008{\natexlab{a}})]{Buffler2008a}
A.~Buffler, et~al., \emph{The Physics Teacher} \textbf{46}, 539
  (2008{\natexlab{a}}).

\bibitem[Kung(2005)]{Kung2005}
R.~L. Kung, \emph{Am. J. Phys.} \textbf{73}, 771--777 (2005).

\bibitem[Redish et~al.(1997)]{Redish1997}
E.~F. Redish, et~al., \emph{Am. J. Phys.} \textbf{65}, 45--54 (1997).

\bibitem[Buffler et~al.(2008{\natexlab{b}})]{Buffler2008}
A.~Buffler, et~al., \emph{Am. J. Phys.} \textbf{76}, 431--437
  (2008{\natexlab{b}}).

\bibitem[Weaver et~al.(2008)]{Weaver2008}
G.~C. Weaver, et~al., \emph{Nat. Chem. Biol.} \textbf{4}, 577--80 (2008).

\bibitem[Horsch et~al.(2012)]{Horsch2012}
E.~Horsch, et~al., \emph{J. Coll. Sci. Teach.} \textbf{41}, 38--43 (2012).

\bibitem[Day and Bonn(2011)]{Day2011}
J.~Day, and D.~Bonn, \emph{Phys. Rev. ST PER} \textbf{7}, 1--14 (2011).

\bibitem[Etkina et~al.(2006)]{Etkina2006b}
E.~Etkina, et~al., \emph{Am. J. Phys.} \textbf{74}, 979 (2006).

\bibitem[Etkina et~al.(2010)]{Etkina2010}
E.~Etkina, et~al., \emph{J. Learn. Sci.} \textbf{19}, 54--98 (2010).

\bibitem[Lederman et~al.(2002)]{Lederman2002}
N.~G. Lederman, et~al., \emph{J. Res. Sci. Teach.} \textbf{39}, 497 (2002).

\bibitem[Redish et~al.(1998)]{Redish1998}
E.~F. Redish, et~al., \emph{Am. J. Phys.} \textbf{66}, 212--224 (1998).

\bibitem[Elby(2001)]{Elby2001}
A.~Elby, \emph{Am. J. Phys.} \textbf{69}, S54 (2001).

\bibitem[Adams et~al.(2006)]{Adams2006}
W.~Adams, et~al., \emph{Phys. Rev. ST PER} \textbf{2}, 1--14 (2006).

\bibitem[web(2012)]{website:E-CLASS1140_Short}
http://tinyurl.com/e-class-survey (2012).

\bibitem[Zwickl et~al.(2011)]{Zwickl2011}
B.~Zwickl, et~al., \emph{PERC Proc.} \textbf{391}, 391--394 (2011).

\bibitem[McCaskey et~al.(2004)]{McCaskey2004}
T.~L. McCaskey, et~al., \emph{PERC Proc.} \textbf{790}, 57--60 (2004).

\bibitem[Gray et~al.(2008)]{Gray2008a}
K.~Gray, et~al., \emph{Phys. Rev. ST PER} \textbf{4}, 1--10 (2008).

\bibitem[Taylor(1997)]{Taylor1997}
J.~R. Taylor, \emph{{An Introduction to Error Analysis}}, University Science
  Books, 1997.

\end{thebibliography}

\IfFileExists{\jobname.bbl}{}
 {\typeout{}
  \typeout{******************************************}
  \typeout{** Please run "bibtex \jobname" to optain}
  \typeout{** the bibliography and then re-run LaTeX}
  \typeout{** twice to fix the references!}
  \typeout{******************************************}
  \typeout{}
 }

\end{document}